\documentclass[
]{ceurart}

\usepackage{listings}
\usepackage{url}
\usepackage{graphicx}
\usepackage{caption}
\usepackage{csvsimple}
\usepackage{multicol}

\begin{document}


\copyrightyear{2023}
\copyrightclause{Copyright © 2023 for this paper by its authors. Use permitted under Creative Commons License Attribution 4.0 International (CC BY 4.0).}

\conference{8th International Workshop on Satisfiability Checking and Symbolic Computation, July 28, 2023, Tromsø, Norway, Collocated with ISSAC 2023}

\title{Data Augmentation for Mathematical Objects}

\author[1]{Tereso del R{\'i}o}[%
orcid=0000-0003-3769-5478,
email=delriot@coventry.ac.uk,
url=https://sites.google.com/view/tereso,
]
\address[1]{Coventry University, UK}

\author[1]{Matthew England}[%
orcid=0000-0001-5729-3420,
email=Matthew.England@coventry.ac.uk,
url=https://matthewengland.coventry.domains,
]

\begin{abstract}
  This paper discusses and evaluates ideas of data balancing and data augmentation in the context of mathematical objects: an important topic for both the symbolic computation and satisfiability checking communities, when they are making use of machine learning techniques to optimise their tools.  We consider a dataset of non-linear polynomial problems and the problem of selecting a variable ordering for cylindrical algebraic decomposition to tackle these with.  By swapping the variable names in already labelled problems, we generate new problem instances that do not require any further labelling when viewing the selection as a classification problem. We find this augmentation increases the accuracy of ML models by 63\% on average. We study what part of this improvement is due to the balancing of the dataset and what is achieved thanks to further increasing the size of the dataset, concluding that both have a very significant effect. We finish the paper by reflecting on how this idea could be applied in other uses of machine learning in mathematics.
\end{abstract}



\maketitle

\section{Introduction}

\subsection{Machine learning and cylindrical algebraic decomposition}

Cylindrical Algebraic Decomposition (CAD) is an algorithm which, given a set of polynomials,  decomposes the space in which they are defined into regions in which they are sign-invariant \cite{Collins1975x}.  CAD has many potential applications, however, its theoretical and practical complexity is doubly exponential \cite{Brown2007}, reducing the scope of its use in practice. In recent years, CAD has been a central component of the collaboration between the Symbolic Computation and Satisfiability Checking communities which meet in this SC$^2$ workshop.  For example, there have been adaptions of CAD for use as an SMT theory solver \cite{Kremer2020}, a repackaging of CAD theory into new algorithms better suited for satisfiability (namely cylindrical algebraic coverings \cite{Abraham2021} and the use of CAD in the model constructing calculus \cite{Jovanovic2012}), and the NuCAD algorithm which uses some of these ideas to tackle more general quantifier elimination problems \cite{Brown2015}. 

CAD requires a declared variable ordering.  In the satisfiability context, the variable ordering is unspecified (any may be chosen to gain a correct result), and in the quantifier elimination context, there is freedom within quantifier blocks (as swapping the order of quantified variables changes the meaning only if the quantifiers are different.  These choices of variable ordering may not affect the correctness of the end result but they can have a huge impact on the resources required by these algorithms.  In fact, Brown and Davenport found in \cite{Brown2007} that there are a family of problems for which in the worst ordering the complexity grows doubly exponentially ($2,4,16,256,4294967296, \dots$) while another ordering has a constant complexity.

Since the community realised the importance of variable ordering, various human-made heuristics have been developed for the choice, e.g. \cite{Brown2004}, \cite{Dolzmann2004}, \cite{Bradford2013}, \cite{Wilson2015}, \cite{boulier_new_2022}.  There have also been some experiments with dynamical variable orderings in the satisfiability context \cite{NKA19}.  None of these heuristics is perfect; all have room for improvement in their choices.  This led to a new strain of research which applied Machine Learning (ML) models to make the choice: first in selecting which human-made heuristic to follow \cite{Huang2014},  and later selecting the ordering directly \cite{Florescu2019a}, \cite{Florescu2019b}, \cite{FE20b}, \cite{Chen2020}.  While these models have demonstrated good performance, there are barriers to their use such as the lack of meaningful training data, and the unbalanced nature of such data that does exist.

This paper proposes to balance and augment the existing datasets by exploiting the arbitrary nature of the variable representations within (the variable names). We note that this idea has been independently proposed recently in the preprint \cite{hester_revisiting_2023}. The present paper makes similar findings on the benefits of augmentation as  \cite{hester_revisiting_2023} and further explores how those benefits split between solving the problem of unbalanced data and increasing the data size.   

\subsection{Data augmentation}

Data augmentation consists of generating new data instances from existing ones. It is a widely-used technique in ML more generally, where the ability to increase the dataset can help tackle over-fitting, and increase the accuracy of the resulting model. Moreover, it can be used to mitigate the biases in the dataset and to reduce the cost of labelling \cite{Shorten2019}.

Data augmentation is commonly used to generate new images in Computer Vision ML applications. Let us take these ideas used in computer vision as an analogy for generating new mathematical objects.  For example, it is clear to any human that a picture of an arrow pointing to the right that is rotated 90 degrees clockwise results gives a picture of an arrow pointing downwards. This can be very useful, imagine that your dataset contains 268 images:  4 images of arrows pointing downwards, 35 pointing left, 56 pointing upwards, and 173 pointing to the right.  
This dataset is very unbalanced, and any model trained on it would likely have a bias towards predicting that the arrow points to the right and against predicting that the arrow points downwards. However, by simply using image rotations the dataset can be balanced to contain 67 images for each of the classes. Furthermore, since you can obtain three extra images from each of the images in the original dataset, we could actually obtain an augmented dataset of 1072 images with 268 of each class.

Returning to our mathematical context, our objects are sets of polynomials (possibly used to form polynomial constraints).  For example, $\{x_1^2-x_2,x_3^3-1\}$.  We can determine, by computing and comparing CADs, that the optimal variable ordering to compute a CAD for this set is $x_2\succ x_1\succ x_3$. Now observe that simply by swapping the names of the variables $x_1$ and $x_2$ we may obtain the new set of polynomials $\{x_2^2-x_1,x_3^3-1\}$, in which we know, without any further CAD computation, that the optimal variable ordering is $x_1\succ x_2\succ x_3$.

\subsection{Plan of the paper}

In this paper, we will use data augmentation to balance our initially unbalanced polynomial dataset, obtaining an improvement in the accuracy of the models. Then, we will see how much more accuracy will be improved by generating the maximum number of instances possible with the data augmentation tools we have.  Section \ref{sec:dataset} outlines our methodology in creating a labelled dataset to use for ML to select a CAD variable ordering and Section \ref{sec:modify} how we have balanced and augmented that dataset.  Then in Section 4, we compare the performance of ML models trained and tested on these various datasets.  We finish in Section \ref{sec:final} with conclusions, a comparison with some similar work in the preprint \cite{hester_revisiting_2023}, and ideas for future work.

The dataset and code used to generate the datasets and results described in this paper can be found on  \href{https://github.coventry.ac.uk/delriot/AugmentingMathematicalDataset}{GitHub} here:
\begin{center}
    \url{https://github.coventry.ac.uk/delriot/AugmentingMathematicalDataset}
\end{center}

\section{Creating a Dataset}
\label{sec:dataset}

There are three steps towards creating a dataset suitable for ML in our context: finding a collection of sets of meaningful polynomials, choosing a methodology to represent each of these sets to an ML model, and then a system for labelling them (identifying the best CAD ordering).  We describe each of these steps in the following subsections.

\subsection{Source of polynomial problems}

The collection of sets of polynomials we use will be those problems in the QF\_NRA collection of the SMT-LIB library \cite{Barrett2016} which involve three variables.  These examples are all satisfiability problems and thus do not represent the full application range of CAD which can also address quantifier elimination.  However, there are no sizeable datasets of QE problems we are aware of.  The problems in the SMT-LIB do mostly emit from real applications making performance upon them meaningful.  Common sources are problems include the theorem prover MetiTarski \citep{Paulson2012}, attempts to prove termination of term-rewrite systems, verification conditions from Keymaera \citep{PQR09}, and curated sets of problems from geometry \cite{BKRVV21},  economics \citep{MDE18} and biology \citep{BDEEGGHKRSW20}.   

\subsection{Representing sets of polynomials}

Representing sets of polynomials for ML is not an easy task.  First, their size can vary: we have already chosen to fix the number of variables but there could then still be an arbitrary number of polynomials, and each of these polynomials can have a great many different terms (although in practice each has not very many). 

To represent a set of polynomials we will follow the methodology of \cite{Florescu2019a} where polynomial sets are represented by a vector of real (floating point) numbered features, with those features generated algorithmically through simple operations generated in turn for each variable.  For example, one feature is the sum across the polynomials of the average of the degree of $x_1$ across the monomials. For the set of polynomials $\{x_2^2-x_2x_1, x_3^3x_1-x_1^2+1\}$, this feature is $3/2$, as the average degree of $x_1$ in the first polynomial is $\tfrac{1}{2}$ and $1$ in the second.  

As well as sum and average, the framework we use can apply the operations of maximum, sum, average, and average of non-zero terms. We also have the possibility of taking the sign at any point. Another example feature is the sum across the polynomials of the sign of the sum of the degree of $x_2$ across the monomials (which simplifies to the number of polynomials that contain the variable $x_2$).  For the previous set of polynomials, this feature is $1$, because the sum of the degree of $x_1$ is $3$ in the first polynomial and $0$ in the second.  Moreover, the degree of the variable can be substituted by $sv_{x_i}$, the total degree of the monomial if the monomial includes such a variable (it is 0 otherwise).  E.g. $sv_{x_1}$ is 4 for the monomial $x_1x_2x_3^2$ and 0 for the monomial $x_2^3x_3$ because $x_1$ does not appear in the latter.  See \cite{Florescu2019a} for further details.

Applying this process results in 384 features to describe a set of polynomials in three variables, of which 195 are essentially distinct (not in a linear relationship with any other feature in our dataset). We thus use these 195 features to represent a set of polynomials in 3 variables.




\subsection{Labelling the sets of polynomials}

In the case of sets of polynomials of three variables, there are six possible variable orderings.  A CAD has been computed in Maple \cite{CM14a} for each ordering for every problem in our dataset, and we timed how long this took, discarding any example in which all orderings timed out (took more than 60 seconds). The label of the set of polynomials is the number associated with the ordering, as given in Table \ref{tab: orderings}, whose CAD required the lowest computation time.  Thus we form a labelled dataset for an ML classification problem.

\begin{table}
\centering
\begin{tabular}{|c|c|}
\hline
\multicolumn{1}{|l|}{\textbf{Ordering Name}} & \textbf{Ordering } \\ \hline
Ordering 0  & $x_1 \succ x_2 \succ x_3$    \\ \hline
Ordering 1  & $x_1 \succ x_3 \succ x_2$    \\ \hline
Ordering 2  & $x_2 \succ x_1 \succ x_3$    \\ \hline
Ordering 3  & $x_2 \succ x_3 \succ x_1$    \\ \hline
Ordering 4  & $x_3 \succ x_1 \succ x_2$    \\ \hline
Ordering 5  & $x_3 \succ x_2 \succ x_1$    \\ \hline
\end{tabular}
\caption{The six possible variable orderings}
\label{tab: orderings}
\end{table}

\section{Modifying the Dataset}
\label{sec:modify}

The dataset described in the previous section has 1019 instances: 406 labelled 0, 93 labelled 1, 135 labelled 2, 51 labelled 3, 202 labelled 4 and 132 labelled 5. There is hence a clear imbalance in this dataset that will likely result in a bias in models  trained upon it.

We split this dataset into an original testing dataset containing 20\% of the instances (815) and an original training dataset containing the rest.

\subsection{Balancing the dataset}

We first randomly changed the label of each instance permuting the variable names in the underlying polynomials. This is done in both of the original datasets (training and testing), obtaining a balanced training dataset and a balanced testing dataset of the same sizes as the original training and testing sets.





\subsection{Augmenting the dataset}

However, nothing is stopping us from adding all of the six possible re-orderings for each problem to the dataset: each would have a different label which we know without any further labelling.  By adding all the possibilities we obtain a perfectly balanced dataset with six times more data than the original one.  The sizes of all these datasets are shown in Table \ref{tab:instances_datasets}.

\begin{table}[th]
    \centering
        \begin{tabular}{|l|l|l|l|l|l|l|l|}\hline%
            \bfseries Dataset & \bfseries 0 & \bfseries 1 & \bfseries 2 & \bfseries 3 & \bfseries 4 & \bfseries 5 & \bfseries Total
            \begin{filecontents*}{Datasets/datasetInstances2.csv}
dataset,zero,one,two,three,four,five,total
,,,,,,,
train unbalanced dataset,326,74,105,41,163,106,815
,,,,,,,
train balanced dataset,126,113,149,138,144,145,815
,,,,,,,
train augmented dataset,815,815,815,815,815,815,4890
,,,,,,,
test unbalanced dataset,80,19,30,10,39,26,204
,,,,,,,
test balanced dataset,31,34,32,38,34,35,204
,,,,,,,
test augmented dataset,204,204,204,204,204,204,1224
\end{filecontents*}
\csvreader[head to column names]{Datasets/datasetInstances2.csv}{}{%

            \ifcsvstrequal{\dataset}{}
            {}
            {
            \\\hline \texttt{\dataset} & \zero & \one & \two & \three & \four & \five & \total
            }
            }
            \\\hline
    \end{tabular}
    \caption{Number of instances of each class that each dataset has. \label{tab:instances_datasets} }
\end{table}

\section{Performance of Models Trained on Different Datasets}
\label{sec:results}


Tables \ref{tab:accuracies_training_datasets_unbalanced}, \ref{tab:accuracies_training_datasets_balanced} and \ref{tab:accuracies_training_datasets_augmented} compare how the ML models trained using the Unbalanced, Balanced and Augmented datasets respectively perform on each of the testing datasets (columns in the tables).  The best performance on each dataset (table column) is highlighted in bold.  Recall that this problem is making classifications from 6 possible variable orderings.  Thus a random classification would on average have an accuracy of 0.17.  We see that all models do better than random, but that there are significant differences in performance.  

We note that our experiments used models from the Python \texttt{sklearn} library \cite{SciKitLearn2011}: K-Nearest Neighbours (KNN), Decision Tree (DT), Support Vector Classifier (SVC), Random Forest (RF), and Multi-Layer Perceptron (MLP).  Each ML model training process followed \cite{Florescu2019a} in first using cross-validation to choose the hyper-parameters of the model.  

We can observe in Table \ref{tab:accuracies_training_datasets_unbalanced} that the models trained with unbalanced data perform very well on unbalanced data. However, when tested in data that is been balanced (the Balanced and Augmented datasets), these models perform terribly, showing that the good results on unbalanced data occur only because both datasets are unbalanced in the same way.   We note that some models were more affected by this than others (e.g. SVC had the biggest drop in performance and RF the least)

Comparing Tables \ref{tab:accuracies_training_datasets_unbalanced} and \ref{tab:accuracies_training_datasets_balanced} it is possible to observe that when testing on data that is balanced, training with a balanced dataset is an asset, in fact, the results improve by 27\% on average.  Performance on the balanced and augmented datasets in universally better.  On the unbalanced dataset, the models trained with balanced data do not perform quite as well as those trained with unbalanced data, but a good deal of the performance is recovered.   

Comparing Tables \ref{tab:accuracies_training_datasets_balanced} and \ref{tab:accuracies_training_datasets_augmented} one observes that fully augmenting the dataset is superior to just balancing it for all models on any of our datasets.

Finally, comparing Tables \ref{tab:accuracies_training_datasets_unbalanced} and \ref{tab:accuracies_training_datasets_augmented} it is possible to observe that when testing on unbalanced data the improvement in performance obtained by augmenting the dataset is similar in scale to that gained by training dataset on a dataset that has the same imbalance as the testing data:  three of the five models perform better on the unbalanced dataset when trained with augmented data and the other two come close.   When comparing performance on a balanced testing dataset is balanced, the improvement from using balanced data or augmented data for training is significant: an increase in 63\% of accuracy on average.

\begin{table}[t]
    \centering
        \begin{tabular}{|l|l|l|l|l|l|}\hline%
            \bfseries Testing dataset & \bfseries Unbalanced & \bfseries Balanced & \bfseries Augmented
            \begin{filecontents*}{Datasets/ml_trained_in_Normal.csv}
Name,Normal,Balanced,Augmented
KNN-Unbalanced,0.51,0.21,0.26
DT-Unbalanced,0.53,0.31,0.31
SVC-Unbalanced,0.48,0.23,0.2
RF-Unbalanced,0.58,0.35,0.37
MLP-Unbalanced,0.51,0.32,0.32
\end{filecontents*}
\csvreader[head to column names]{Datasets/ml_trained_in_Normal.csv}{}{%
            \ifcsvstrequal{\Name}{RF-Unbalanced}
            {
            \\\hline \texttt{\Name} & \textbf{\Normal} & \textbf{\Balanced} & \textbf{\Augmented}
            }
            {
            \ifcsvstrequal{\Name}{KNN}
            {
            \\\hline \texttt{\Name} & \Normal & \Balanced & \textbf{\Augmented}
            }
            {
            \\\hline \texttt{\Name} & \Normal & \Balanced & \Augmented
            }
            }
            }
            \\\hline
    \end{tabular}
    \caption{Accuracy of models trained on the unbalanced dataset, when tested on the different testing datasets. \label{tab:accuracies_training_datasets_unbalanced} }
\end{table}

\begin{table}[t]
    \centering
        \begin{tabular}{|l|l|l|l|l|l|}\hline%
            \bfseries Testing dataset & \bfseries Unbalanced & \bfseries Balanced & \bfseries Augmented
            \begin{filecontents*}{Datasets/ml_trained_in_Balanced.csv}
Name,Normal,Balanced,Augmented

KNN-Balanced,0.41,0.36,0.38

DT-Balanced,0.43,0.45,0.45

SVC-Balanced,0.25,0.3,0.28

RF-Balanced,0.49,0.52,0.52

MLP-Balanced,0.45,0.43,0.44

\end{filecontents*}
\csvreader[head to column names]{Datasets/ml_trained_in_Balanced.csv}{}{%
            \ifcsvstrequal{\Name}{RF-Balanced}
            {
            \\\hline \texttt{\Name} & \textbf{\Normal} & \textbf{\Balanced} & \textbf{\Augmented}
            }
            {
            \ifcsvstrequal{\Name}{KNN}
            {
            \\\hline \texttt{\Name} & \Normal & \Balanced & \textbf{\Augmented}
            }
            {
            \\\hline \texttt{\Name} & \Normal & \Balanced & \Augmented
            }
            }
            }
            \\\hline
    \end{tabular}
    \caption{Accuracy of models trained on the balanced dataset, when tested on the different testing datasets. \label{tab:accuracies_training_datasets_balanced} }
\end{table}

\begin{table}[t]
    \centering
        \begin{tabular}{|l|l|l|l|l|l|}\hline%
            \bfseries Testing dataset & \bfseries Unbalanced & \bfseries Balanced & \bfseries Augmented
            \begin{filecontents*}{Datasets/ml_trained_in_Augmented.csv}
Name,Normal,Balanced,Augmented

KNN-Augmented,0.54,0.55,0.53

DT-Augmented,0.54,0.55,0.54

SVC-Augmented,0.46,0.48,0.49

RF-Augmented,0.62,0.63,0.61

MLP-Augmented,0.48,0.5,0.51

\end{filecontents*}
\csvreader[head to column names]{Datasets/ml_trained_in_Augmented.csv}{}{%
            \ifcsvstrequal{\Name}{RF-Augmented}
            {
            \\\hline \texttt{\Name} & \textbf{\Normal} & \textbf{\Balanced} & \textbf{\Augmented}
            }
            {
            \ifcsvstrequal{\Name}{KNN}
            {
            \\\hline \texttt{\Name} & \Normal & \Balanced & \textbf{\Augmented}
            }
            {
            \\\hline \texttt{\Name} & \Normal & \Balanced & \Augmented
            }
            }
            }
            \\\hline
    \end{tabular}
    \caption{Accuracy of models trained on the augmented dataset, when tested on the different testing datasets. \label{tab:accuracies_training_datasets_augmented} }
\end{table}

\section{Final Thoughts}
\label{sec:final}

\subsection{Conclusions}
\label{sec:conc}

Our first conclusion is that, for this problem, training on an unbalanced dataset does indeed lead to overfitting and poor performance when the models are utilised on a balanced dataset.  The performance in the Unbalanced column in Table \ref{tab:accuracies_training_datasets_balanced}) is much worse than previous reports on such ML models, e.g. \cite{England2019}, and demonstrates the importance of taking note and care of these balance issues.  In general, imbalance is not always inappropriate for ML:  some applications will naturally have imbalanced data and the ML models should be aware of this.  However, in our case, we seek heuristics for choosing variable orderings for CAD applied in general and there is little rationale to suppose general CAD applications favour one ordering over another\footnote{except perhaps the existence of the SMT-LIB data!}.  Thus our advice is to ensure ML models for such applications are trained on balanced data. 

Our second conclusion is that a good deal of the ML performance can be recovered by simply training on balanced data, re-validating the value of the data-science-led approach to this task that those original papers posited.  

Our third conclusion is that it is beneficial to go further and use maximum data augmentation:  all models benefitted from this over just balancing the data no matter which dataset they are tested on.  Using a balanced dataset instead of an unbalanced one of the same size allowed the accuracy of the models to improve on average by 27\%. But using a dataset fully augmented to thus multiply the size by six allowed the accuracy of the models to improve on average by 63\%.  In fact, the performance lost from the original unbalanced case is basically recovered this way.  

Finally, we note that these ideas should generalise easily to variable ordering choice for the other decision procedures of non-linear real arithmetic commonly found in the wider toolchains of the SC$^2$ community.

\subsection{Comparison with the work of Hester et al. (2023)}
\label{sec:Hester}

Let us now compare the results on this paper with the ones obtained in the recent preprint \cite{hester_revisiting_2023}. Table 2 in \cite{hester_revisiting_2023} presents the accuracies of trained models on different datasets: note that both their `Training Set 2' and `Dataset 1' contain instances in which the models have been trained, meaning that `Testing Set 2' is the most appropriate column for evaluation in that table. That column shows similar results to the ones shown in this paper.

We note that the original dataset in \cite{hester_revisiting_2023} contained 6895 instances while in our paper the initial dataset only contained 1019 instances. This is because, even though both datasets have the same ultimate source (the SMT-LIB), our dataset had been stripped of duplicate instances (those problem instances whose CAD tree structure is identical for every variable ordering), as described in detail in Section 4.1 of  \cite{boulier_new_2022}.  We view this as a necessary step to meaningful use of the QF\_NRA section of the SMT-LIB where there are many \emph{very} similar problems.

This comparison with \cite{hester_revisiting_2023} shows that the size alone of the dataset is not what matters (since \cite{hester_revisiting_2023} has similar accuracy to the models presented here despite training with much more data).  Rather it is the number of qualitatively different problems within the dataset. I.e. there is little benefit to including multiple very similar problems.  It may seem that data augmentation adds no new information, but since the ML models are not aware of these symmetries by exposing them with augmentation we actually give them access to this information.

\subsection{Future work}
\label{sec:Future}

Given the success of this data augmentation, an obvious area for future work is to look for additional augmentation techniques.  Returning to the computer vision analogy: rotations are not the only augmentation tool, there are others also e.g. mirror reflections. Regarding mathematical objects, a corresponding augmentation technique may be substituting a variable with its negative, which would create a new instance without the need for any further labelling. We could also consider more involved variable transformations, however, these would most likely require additional CAD computations for data labelling, which is the most expensive part of this whole process.  

We note that these ideas of data augmentation could be generalised to other mathematical object datasets.  One should reflect on which parts of the representation of a mathematical object are arbitrary to the problem at hand.  For example, in  \cite{lample_deep_2020} the authors consider symbolic integration by ML, with mathematical expressions represented as natural text.  The order of the operands in commutative operations is arbitrary (e.g. $x\wedge 2+y*z$ is the same expression as $z*y+x\wedge 2$). This could be exploited to generate an exorbitant amount of new instances that do not require any relabelling!

\begin{acknowledgments}
TdR is supported by Coventry University and a travel grant from the London Mathematical Society (LMS).  ME is supported by UKRI EPSRC Grant EP/T015748/1, \emph{Pushing Back the Doubly-Exponential Wall of Cylindrical Algebraic Decomposition} (the DEWCAD Project).
\end{acknowledgments}


\bibliography{library}



\end{document}
